\documentclass[a4paper,fleqn,usenatbib]{mnras}

\usepackage{newtxtext,newtxmath}
\usepackage[T1]{fontenc}
\usepackage{ae,aecompl}

\usepackage{graphicx}	% Including figure files
\usepackage{amsmath}	% Advanced maths commands
\usepackage{amssymb}	% Extra maths symbols
\usepackage{xspace}        %  For problem with missing space in using newcommand

\title[Radio structure in PDS~456]{A radio structure resolved at the deca-parsec scale in radio-quiet quasar PDS 456 with an extremely powerful X-ray outflow}

\author[J. Yang et al.]
{Jun Yang,$^{1,2}$\thanks{E-mail: jun.yang@chalmers.se}
Tao An,$^{2}$\thanks{E-mail: antao@shao.ac.cn}
Fang Zheng,$^{2,3}$
Willem A. Baan,$^{4, 5}$
Zsolt Paragi,$^{6}$ 
\newauthor
Prashanth Mohan,$^{2}$
Zhongli Zhang,$^{2}$
and Xiang Liu$^{4}$
\\
\\
$^{1}$Department of Space, Earth and Environment, Chalmers University of Technology, Onsala Space Observatory \\
SE-439 92 Onsala, Sweden \\
$^{2}$Shanghai Astronomical Observatory, Key Laboratory of Radio Astronomy, Chinese Academy of Sciences \\
200030 Shanghai, P.R. China \\
$^{3}$University of Chinese Academy of Sciences, 19A Yuquan Road, Shijingshan District, 100049 Beijing, P.R. China \\
$^{4}$Xinjiang Astronomical Observatory, Key Laboratory of Radio Astronomy, Chinese Academy of Sciences \\
150 Science 1-Street, 830011 Urumqi, P.R. China \\
$^{5}$Netherlands Institute for Radio Astronomy ASTRON, NL-7991 PD Dwingeloo, the Netherlands \\
$^{6}$Joint Institute for VLBI ERIC (JIVE), Postbus 2, NL-7990 AA Dwingeloo, the Netherlands \\
}
\date{Accepted 2018 XXX. Received 2018 YYY; in original form 2018 ZZZ}

\pubyear{2018}

\begin{document}
\label{firstpage}
\pagerange{\pageref{firstpage}--\pageref{lastpage}}
\maketitle

% <=250 words
\begin{abstract}
Active galactic nuclei (AGN) accreting at rates close to the Eddington limit can host radiatively driven mildly relativistic outflows. Some of these X-ray absorbing but powerful outflows may produce strong shocks resulting in a significant non-thermal emission. This outflow-driven radio emission may be detectable in the radio-quiet quasar PDS~456 since it has a bolometric luminosity reaching the Eddington limit and a relativistic wide-aperture X-ray outflow with a kinetic power high enough to quench the star formation in its host galaxy. To investigate this possibility, we performed very-long-baseline interferometric (VLBI) observations of the quasar with the European VLBI Network (EVN) at 5 GHz. The EVN image with the full resolution reveals two faint and diffuse radio components with a projected separation of about 20 pc and an average brightness temperature of around two million Kelvin. In relation to the optical sub-mas-accuracy position measured by the \textit{Gaia} mission, the two components are very likely on opposite sides of an undetected radio core. The VLBI structure at the deca-pc scale can thus be either a young jet or a bidirectional radio-emitting outflow, launched in the vicinity of a strongly accreting central engine. Two diffuse components at the hecto-pc scale, likely the relic radio emission from the past AGN activity, are tentatively detected on each side in the low-resolution EVN image.   
\end{abstract}

\begin{keywords}
galaxies: active -- quasars: individual: PDS~456 -- radio continuum: galaxies
\end{keywords}

%%%%%%%%%%%%%%%%%%%%%%%%%%%%%%%%%%%%%%%%%%%%%%%%%%

%%%%%%%%%%%%%%%%% BODY OF PAPER %%%%%%%%%%%%%%%%%%

\section{Introduction}
\label{sec1}
% Eddington ratios and their importance
The accretion rate can regulate the radio emission in active galactic nuclei (AGN) either directly in the form of wide-angle mildly relativistic winds \citep[e.g.,][]{Yuan2014} or by supplementing the jet synchrotron emission. Statistical studies of AGN indicate an anti-correlation between the accretion rate and the radio loudness \citep[e.g.,][]{Greene2006, Panessa2007, Sikora2007}. For accretion rates approaching the Eddington limit, AGN tend to be radio quiet \citep[e.g., ][]{Greene2006}, and can host powerful outflows, which can quench star-formation activity \citep[e.g.,][]{Nardini2015, Tombesi2015, Tombesi2016} and address the supermassive black hole - host galaxy coevolution through feedback \citep[e.g.,][]{Kormendy2013}. Some high-speed ($\sim$0.1~c) outflows may produce strong interactions or shocks and thus may create radio-emitting sources \citep[e.g.,][]{Nims2015, Zakamska2014}. Very-long-baseline interferometric (VLBI) observations of optically luminous radio-quiet quasars will allow us to search for radio-emitting wide-angle but powerful outflows to provide direct evidence for this mode of AGN feedback \citep[e.g,][]{Giroletti2017, Wylezalek2018}.  

PDS~456 is a radio-quiet quasar at $z=0.184$ \citep{Simpson1999} with a black hole mass of $M_\mathrm{bh}\sim10^9$~M$_{\sun}$ \citep{Nardini2015} and is one of a few AGN known to host an extremely powerful outflow \citep[$\sim$10$^{46}$~erg s$^{-1}$,][]{Nardini2015, Tombesi2016} and a compact radio counterpart \citep{Yun2004}. It is one of the most luminous quasars in the Universe at $z<0.3$ \citep{Simpson1999, Torres1997} and has a bolometric luminosity of $L_\mathrm{bol}\sim10^{47}$ erg\,s$^{-1}$ \citep{Reeves2000}, close to its Eddington luminosity $L_\mathrm{Edd}$ \citep{Nardini2015}. X-ray spectroscopic observations indicate a powerful mildly relativistic outflow covering a solid angle of 3$\pi$ and active over time scales of years \citep{Matzeu2017, Nardini2015}. The X-ray outflow shows a tight correlation with the X-ray luminosity, indicating that the outflow is likely driven by the radiation pressure \citep{Matzeu2017}. The radio counterpart of PDS~456 has an optically thin spectrum with relatively high flux densities: $24\pm5$~mJy at 1.2~GHz and 4.5$\pm$0.9~mJy at 8.4 GHz, and a spectral index of $-0.85\pm0.10$,  and shows a point-source structure in the Very Large Array (VLA) images \citep{Yun2004}.  As the radio continuum luminosity is nearly an order of magnitude higher than the estimated far-infrared luminosity, AGN activity was inferred to dominate the source radio emission \citep{Yun2004}. To investigate the origin of the radio emission and association with AGN activity, we performed very high-resolution imaging observations of PDS~456 with the European VLBI Network (EVN) at 5 GHz. 

This work is organised as follows. Section~\ref{sec2} describes the EVN experiment and the data reduction. Section~\ref{sec3} presents the source radio morphology on sub-kpc scales. Section~\ref{sec4} discusses the radio core, the AGN activity and the origin of the radio VLBI structure, with concluding remarks in Section~\ref{sec5}. Throughout the paper, a standard $\Lambda$CDM cosmological model with H$_\mathrm{0}$~=~71~km~s$^{-1}$~Mpc$^{-1}$, $\Omega_\mathrm{m}$~=~0.27,  $\Omega_{\Lambda}$~=~0.73 is adopted; the images then have a scale of 3.1~pc mas$^{-1}$.

\section{Observations and data reduction}
\label{sec2}

We observed PDS 456 with the e-EVN at 5~GHz on 2016 February 3. During the observations, each station had a data transferring speed of 1024~Mbps (16 subbands in dual polarisation, 16 MHz per subband, 2-bit quantization). The correlation was done in real-time mode by the EVN software correlator \citep[SFXC,][]{Keimpema2015} at JIVE (Joint Institute for VLBI, ERIC) using 2-s integration time and 32 frequency points per subband. The participating EVN stations included Effelsberg, Medicina, Noto, Onsala, Torun, Yebes, Westerbork (single antenna) and Hartebeesthoek.

The e-EVN observations of PDS 456 were performed with the phase-referencing technique. The compact source J1724$-$1443  (about 59 arcmin apart) was employed as the phase-referencing calibrator. Its position is R.A.~$=17^{\rm h}24^{\rm m}46\fs96656$, Dec.~$=-14\degr43\arcmin59\farcs7610$ (J2000) in the source catalogue from the Goddard Space Flight Centre VLBI group. In the existing geodetic VLBI observations, the calibrator had a correlation amplitude of $\ga$0.15~Jy at 8.4~GHz on the long baselines. The correlation position for PDS~456 is R.A.=$17^{\rm h}28^{\rm m}19\fs7920$, Dec.=$-14\degr15\arcmin55\farcs912$ (J2000). The nodding observations used a cycle period of about four minutes (1.5~min for J1724$-$1443, 2.5~min for PDS~456) and lasted a total of about two hours. All telescopes had an elevation of $\geq$17~deg during the observations. Besides the pair of sources, a bright flat-spectrum radio quasar NRAO~530 \citep[e.g.,][]{An2013}  was also observed to enable the determination of instrumental bandpass shapes.  

The data were calibrated using the National Radio Astronomy Observatory (NRAO) software package Astronomical Image Processing System \citep[\textsc{aips}, ][]{Greisen2003}.  \textit{A-priori} amplitude calibration was performed with the system temperatures and the antenna gain curves. The ionospheric dispersive delays were corrected according to a map of total electron content provided by Global Positioning System (GPS) satellite observations. Phase errors due to antenna parallactic angle variations were removed. A manual phase calibration was carried out with a two-minute scan of NRAO 530 data.  After the instrumental phase delays were removed, the global fringe-fitting and the bandpass calibration were performed.

We first imaged the phase-referencing source J1724$-$1443 using a loop of model fitting and self-calibration in \textsc{difmap} \citep{Shepherd1994} and then re-ran fringe-fitting to remove its structure-dependent phase errors in \textsc{aips} in the second iteration. The calibrator J1724$-$1443 has a single-side core-jet structure with a total flux density of 0.37$\pm$0.02~Jy. Its radio core, i.e. the jet base, is a point source with a flux density of 0.16$\pm$0.01~Jy and its position was used as the reference point in the phase-referencing calibration. We also ran amplitude and phase self-calibrations on the data of J1724$-$1443  and transferred the solutions to the data of PDS~456.

Owing to the limited $uv$ coverage in particular on the long baselines, the deconvolution was performed by fitting the visibility data of PDS~456 directly to some circular Gaussian models in \textsc{difmap} to minimise the potential deconvolution errors of \textsc{clean}. The model-fitting results are reported in Table~\ref{tab1}.  

\begin{table*}
\caption{Circular Gaussian model-fitting results of the radio components detected in PDS~456 with the e-EVN observations at 5 GHz. Columns give (1) component name, (2) Right Ascension, (3) Declination, (4) positional error (including an error of 0.24~mas from the reference source), (5) peak brightness, (6) integrated flux density, (7) FWHM, (8) average brightness temperature, (9) fraction with respect to the VLA total flux density measurement at 5~GHz \citep{Reeves2000} and radio luminosity.    }
\label{tab1}
\begin{tabular}{cccccccccc}
\hline\hline
Component  &                R.A.                                       &                  Dec.                                & $\sigma_{\rm p}$   
                                                                                                                                                                &  $S_{\rm peak}$        & $S_{\rm int}$       & $\theta_{\rm size}$       &     $T_{\rm b}$        & $S_{\rm evn}/S_{\rm vla}$ 
                                                                                                                                                                                                                                                                                                                               & $L_{\rm r}$ \\
             &              (J2000)                                   &                (J2000)                               & (mas)           &   (mJy beam$^{-1}$)  & (mJy)                  &   (mas)                          &     ($10^{\rm 6}$ K) &        & (10$^{39}$ erg s$^{-1}$) \\
\hline
C1         &  $17^{\rm h}28^{\rm m}19\fs78967$  &  $-14\degr15\arcmin55\farcs8550$    &  0.4              &   0.36$\pm$0.03         &  1.09      &     6.9                            &  1.4                & 0.13 & 5.5 \\
C2         &  $17^{\rm h}28^{\rm m}19\fs78943$ &  $-14\degr15\arcmin55\farcs8499$     &  0.6              &   0.18$\pm$0.03         &  0.29      &    2.6                            &  2.6                 &  0.04 & 1.7 \\
X1         &  $17^{\rm h}28^{\rm m}19\fs79224$  &  $-14\degr15\arcmin55\farcs9544$     &  1.9             &    0.47$\pm$0.07        &  0.78       &    15.7                          &   0.2              &  0.09 & 3.8 \\
X2        &  $17^{\rm h}28^{\rm m}19\fs78523$  &  $-14\degr15\arcmin55\farcs8030$     &  2.2             &    0.40$\pm$0.07        &  0.48       &    6.5                            &   0.7               &  0.06 & 2.5 \\

\hline
\end{tabular}
\end{table*}

\section{VLBI imaging results of PDS~456}
\label{sec3}

\begin{figure*}
\centering
\includegraphics[width=0.98\textwidth]{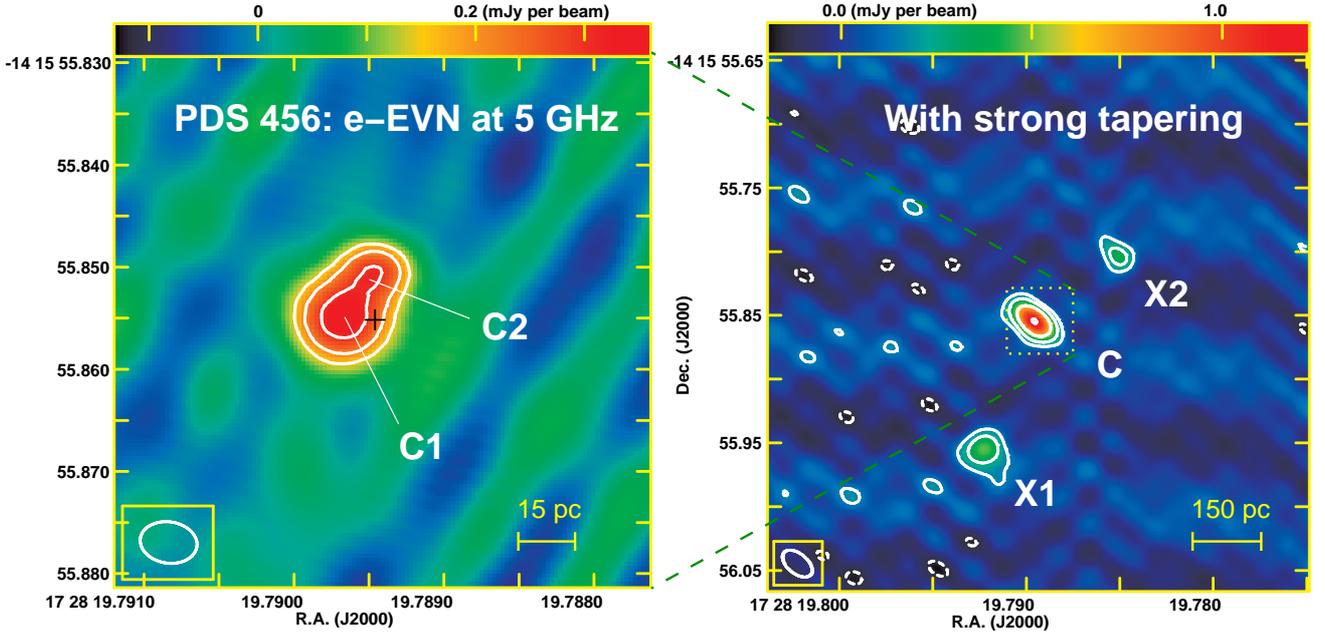}  \\
\caption{
The \textsc{clean} images of the optically luminous radio-quiet quasar PDS 456 observed with the e-EVN at 5-GHz. Left: firm detections of two faint and diffuse components C1 and C2 with natural weighting and slightly Gaussian tapering (0.2 at the $uv$ radius 120 M$\lambda$). The black cross marks the sub-mas-accuracy optical centroid measured by \textit{Gaia}. The full width at half maximum (FWHM) of the synthesised beam is 5.5~mas~$\times$~4.4~mas at 81.2 deg. The contours are 0.09~$\times$~($-$1, 1, 1.4, 2)~mJy~beam$^{-1}$.  The image peak is 0.36 mJy~beam$^{-1}$. Right: tentative detections of two more diffuse radio components X1 and X2 on sub-kpc scales with natural weighting and strong Gaussian tapering (0.5 at the $uv$ radius 5 M$\lambda$). The beam FWHM is 27.1~mas~$\times$~15.5~mas at 50.1 deg. The contours are 0.15~$\times$~($-$1, 1, 2, 4, 8)~mJy~beam$^{-1}$. The image peak is 1.23~mJy~beam$^{-1}$. All the characteristic parameters related to these components are summarised in Table~\ref{tab1}. The first contours in both images are at 3$\sigma$ level. }
\label{fig1}
\end{figure*}

The \textsc{clean} maps of PDS~456 are shown in Fig.~\ref{fig1}.  There are two faint radio components detected with a projected distance of 19~pc and a relative position angle of 146$\degr$, marked as C1 and C2 in the left panel. Two more components, X1 and X2, are tentatively detected at the much larger scales as shown in the right panel.  

The two-component structure in component C is present even by studying dirty maps, i.e. without running de-convolution. Compared to the dirty beam, the noise pattern convolved with component C is much smoother and wider along the extension direction in the initial dirty map with natural weighting. Component C is clearly seen with a peak brightness: 0.64 mJy~beam$^{-1}$ (SNR $\sim$16) when all the baselines within Europe ($\leq33$~M$\lambda$, in which $\lambda$ is wavelength) are included, but is not observed up to an upper limit of  0.3~mJy~beam$^{-1}$ (3$\sigma$) using the long baselines of $\geq$103~M$\lambda$ between the European stations and Hartebeesthoek. Due to this significant decrease in peak brightness, a single point-source structure can be firmly excluded. 

Assuming no significant variability, the flux density of component C represents only 16 per cent of that measured with the VLA \citep[8.2~mJy at 5~GHz,][]{Reeves2000}. Employing a strong Gaussian tapering (e.g., 0.5 at 5~M$\lambda$), two additional diffuse components X1 and X2 at a projected separation of about 550~pc are marginally detected.  As the detections are mainly dependent on the shortest baseline (Effelsberg--Westerbork), it is hard to properly image and locate these faint extended structures. Including these two components, about 68 per cent of the VLA flux density is still unaccounted for in our VLBI image. As a portion of this missing emission is likely associated with these VLBI components, both the total flux density and the size reported in Table~\ref{tab1} are considered as lower limits.    

All the components have relatively low average brightness temperatures, $T_{\rm b}\la10^6$ K.  The last column in Table~\ref{tab1} reports $T_{\rm b}$ , estimated as \citep[e.g.,][]{Condon1982},
\begin{equation}
T_\mathrm{b} = 1.22\times10^{9}\frac{S_\mathrm{int}}{\nu_\mathrm{obs}^2\theta_\mathrm{size}^2}(1+z),
\label{eq1}
\end{equation}
where $S_\mathrm{int}$ is the integrated flux density (in mJy),  $\nu_\mathrm{obs}$ is the observation frequency (in GHz), $\theta_\mathrm{size}$ is the FWHM of the circular Gaussian model (in mas), and $z$ is the redshift.

\section{Discussion}
\label{sec4}

\subsection{Radio AGN activity}
\label{sec4-1}

All VLBI-detected components are located within the optical nuclear region of PDS 456. The centroid of its optical emission (J2000, R.A.$=17^{\rm h}28^{\rm m}19\fs789380$, Dec.$=-14\degr15\arcmin55\farcs85543$, $\sigma_{\rm p}=0.04$~mas), published by the second data release \citep[DR2,][]{Gaia2018} of the \textit{Gaia} mission \citep{Gaia2016}, is marked as a black cross in the left panel of Fig.~\ref{fig1}. As the residuals from the point-source model fitting statistically agree with the assumed observational noise, the astrometric excess noise for PDS 456 is insignificant (i.e., 0 mas) in the \textit{Gaia} DR2 catalogue. The extremely compact optical morphology then allows us to take the optical centroid as a robust marker of the central supermassive black hole hosted in the optical nucleus. The components C1 and C2 offsets of 4.2$\pm$0.4~mas and 5.5$\pm$0.6~mas respectively, with respect to the optical centroid. 

As the resolved components C1 and C2 have significant offsets from the optical centroid, they may not be associated with the radio core, i.e. the base of a well-collimated (mildly) relativistic jet with a flat spectrum. If components C1 and C2 are a pair of episodic ejecta or a radio-emitting wind-like outflow, driven by the central engine, the radio core would be expected at a position near the centre of the pair of components. Assuming that the radio core has a point-source structure, we can set a $3\sigma$ upper limit of 0.3~mJy on the total flux density of the hidden radio core with the data on the long baselines to Hartebeesthoe only. This gives a radio luminosity of $L_{\rm r}=\nu L_\nu\leq1.3\times 10^{39}$~erg~s$^{-1}$.

While there is no evidence for a compact and partially self-absorbed radio core, there seems to be ample evidence for AGN activity as a dominant source of the faint radio structure on sub-kpc scales in PDS 456. All the VLBI components in Fig.~\ref{fig1} are located roughly along a line in northwest-southeast direction. This structural alignment can be naturally formed by AGN activity. It is difficult to associate these radio components with supernovae remnants produced by star-formation activity. A supernova can have a peak monochromatic luminosity up to $L_\nu$ $\sim$10$^{29}$~erg~s$^{-1}$~Hz$^{-1}$ at 5 GHz \citep{Weiler2002}. While, these VLBI components have monochromatic luminosities $L_\nu$ $\geq2.7\times10^{29}$~erg~s$^{-1}$~Hz$^{-1}$ at 5~GHz and extended structures, it is certainly hard to attribute any component to a single young supernova. Additionally, since PDS 456 is not a starburst galaxy like Arp 220 \citep{Yun2004}, we do not expect to observe a great number of  supernova remnants in the nuclear region. 

\begin{figure}
\centering
\includegraphics[width=0.98\columnwidth]{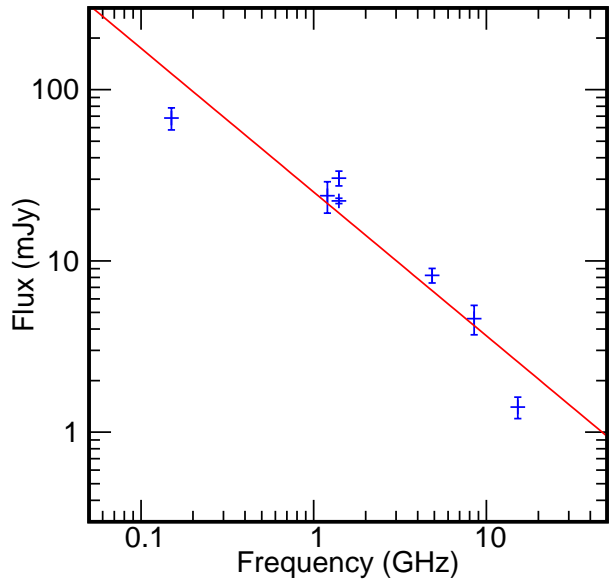}  \\
\caption{
The radio spectrum of PDS 456 observed from 150 MHz to 15 GHz. The red line represents the simple model of $S_\nu=S_{\rm 0}\nu^{\alpha}$, where $S_0=25\pm4$  mJy and $\alpha=-0.84\pm0.11$. }
\label{fig2}
\end{figure}

The radio AGN activity in PDS 456 is relatively recent. The outer components X1 and X2 represent the largest structure at a projected separation of about 0.55 kpc and likely constitute bright relic radio emission from past AGN activity. Assuming a constant apparent expansion speed of 0.2~$c$ \citep[e.g.,][]{Owsianik1998} for young radio sources \citep[typically $\la$10$^6$ yr, e.g., ][]{Fanti1995}, they have an age of $10^4$ yr.  According to the correlation relation between the linear size and the turnover frequency observed in young and bright radio sources \citep[][]{ODea1998}, the observed linear size allow us to predict a turnover at a frequency of about 1 GHz for PDS 456. Fig.~\ref{fig2} displays the radio spectrum of PDS~456 between 0.15 and 15~GHz. Besides the flux densities reported by \citet{Yun2004}, we added two more measurements. The GMRT observations gave a total flux density of 68$\pm$10~mJy at 150 MHz \citep{Intema2017}.  According to the existing VLA archive data (project: AL501) calibrated by the VLA pipeline, a 15 GHz flux density of 1.4$\pm$0.1~mJy was measured. The red line in Fig.~\ref{fig2} is a power-law fit to the data with a best-fit spectral index of $-$0.84$\pm$0.11. The source does not show a distinct turnover at frequencies $\ge 0.15$ GHz.  The relatively steep spectral index and an integrated flux density dominated by VLBI-undetected diffuse structures indicate a paused or slow expansion from early on; these are then in a dying or decaying state.

\subsection{Component C: outflow-driven vs jet-driven}
Components C1 and C2 may originate from a bidirectional radio-emitting outflow in view of their diffuse and faint radio structure. A high-speed, wide-opening-angle and persistent outflow has been indirectly observed in multiple X-ray spectral observations \citep[e.g.][]{Matzeu2017, Nardini2015, Reeves2018} with reports of multiple velocity components, up to 0.46~c \citep{Reeves2018}. An ultraviolet outflow at 0.3~c has also been recently reported by \cite{Hamann2018}. From a sample of 568 luminous quasars, the study of \cite{Zakamska2014} find a strong association between powerful outflows and the radio luminosity in radio quiet AGN; this is indicative of the radio emission originating from shocks driven by the outflow with a conversion efficiency of $\sim 10^{-4}$ between the radio luminosity and the outflow kinetic power. Component C has a radio luminosity of $L_{\rm r}=7\times10^{39}$ erg~s$^{-1}$, a factor of 3$\times$10$^{-7}$ lower than the kinetic power of the X-ray outflow \citep[$2\times10^{46}$ erg~s$^{-1}$, ][]{Nardini2015}. If the radio emission is indeed wind-like outflow-driven, the conversion efficiency is significantly lower than the typical $10^{-4}$. The X-ray outflow then has sufficient kinetic power to drive shocks which produce the observed radio emission. 

The VLBI morphology resembles compact symmetric objects (CSOs) characterized by two-sided mini-lobes or hot spots at the sub-kpc scales \citep[e.g.,][]{Wilkinson1994, Readhead1996, Xiang2005}. PDS 456 may then be identified as a candidate CSO. Owing to the interaction between jets and the surrounding interstellar medium, CSOs usually have mildly-relativistic speeds \citep[e.g.,][]{Owsianik1998, Polatidis2003, An2012a}. Assuming an intrinsically symmetric ejection and a spectral index of 0.85, i.e. the same as the VLA total flux density measurement, the observed flux ratio between components C1 and C2 implies that $\beta\cos\theta_{\rm v}=0.17$~c, where $\beta$ is the intrinsic jet speed and $\theta_{\rm v}$ is the viewing angle \citep[e.g.,][]{Bottcher2012}. If the jet is directed close to the line of sight, e.g. $\theta_{\rm v}\la8\degr$ \citep{Yun2004}, this still gives an intrinsically low jet speed ($\beta\sim\beta\cos\theta_{\rm v}$). \citet{Sokolovsky2011} reported $T_\mathrm{b}\ga10^{8}$~K for a sample of 64 candidate CSOs with much higher radio power. Compared to the sample, the two components have a brightness temperature two orders of magnitude lower and a significantly expanded structure, thus indicating that they likely constitute a pair of young but dying ejecta. Amongst samples of symmetric radio sources with known radio luminosities \citep[e.g.][]{Kunert2010, An2012b, Kunert2016}, they have a relatively low radio power and may fail to significantly expand in size. 

Compared to a scenario involving collimated jets, the expectation of outflow-driven radio emission should enable the observation of a biconical radio structure near the supermassive black hole with a much wider opening angle, e.g., $\sim$100 degrees based on modelling the P-Cygni profile in the broad-band X-ray spectrum of PDS 456 \citep{Nardini2015}. Our high-resolution VLBI image tentatively supports this wide-opening-angle outflow scenario. Assuming that the radio emission at the deca-pc scale comes from strong shocks of the outflow and the outflow driver is centrally located between C1 and C2, a rough estimate of the opening angle is $\sim$130 degree based on component C1 and $\sim$50 degree from component C2. The intrinsic opening angle may be much smaller than the observed opening angle because of the projection effect. To get a reliable estimate of the opening angle, it is necessary for additional epochs of VLBI observations owing to our size estimates being inaccurate in the direction vertical to the outflow axis due to the limited $uv$ coverage and the usage of a circular Gaussian model. 

Regardless of which scenario is likely in PDS~456, our results clearly establish the association between radio emission and AGN activity. Due to the diffuse two-sided faint structure, there is no strong Doppler beaming effect. In case of the scenario of the outflow-driven radio emission, PDS~456 would be the most promising source for a future observational confirmation amongst radio-quiet quasars. To date, clearly positive cases for the scenario have not been identified \citep[e.g.,][]{Wylezalek2018}. In the only previously known candidate, IRAS 17020$+$4544, the non-thermal radio emission and the ultra-fast X-ray outflow are both powered by the central engine \citep[][]{Giroletti2017}. In comparison, here the powerful X-ray outflow is expected to be the driver of the nuclear radio emission. In the alternate scenario, i.e. jet driven emission, PDF~456 would then be a rare example of intermittent and low-power jet activity amongst sources accreting at rates close to the Eddington limit \citep[e.g.][]{Greene2006}.

\section{Conclusions}
\label{sec5}
In order to understand the origin of compact radio emission from the nuclear region, we observed the optically and X-ray luminous but radio-quiet quasar PDS~456 with the e-EVN at 5 GHz. In the image with the full resolution, we found two faint and diffuse radio components with a separation of about 20 pc and an average brightness temperature of around two million Kelvin. In relation to the sub-mas-accuracy \textit{Gaia} position, the two components are putatively located on opposite sides of an unobserved radio core. The radio structure indicates recent jet-driven AGN activity at a low radio power. The new VLBI observations also indicate that the radio emission likely originates from the shocks produced by the X-ray outflows when interacting with the surrounding external medium. Additional sub-kpc scale diffuse components are tentatively detected in the low-resolution image and appear to be relic radio emission from earlier AGN activity. Future VLBI imaging observations can be employed to provide more conclusive evidence to support the outflow-driven scenario and hence, clarify the origin of radio emission in this source.

\section*{Acknowledgements}
This work was partly supported by the SKA pre-research funding from the Ministry of Science and Technology of China (2018YFA0404600) and the Chinese Academy of Sciences (CAS, No. 114231KYSB20170003).  TA thanks the grant supported by the Youth Innovation Promotion Association of CAS. 
% EVN
The European VLBI Network is a joint facility of independent European, African, Asian, and North American radio astronomy institutes. Scientific results from data presented in this publication are derived from the following EVN project code(s): EY024A. 
% VLA
The National Radio Astronomy Observatory is a facility of the National Science Foundation operated under cooperative agreement by Associated Universities, Inc.
% Gaia DR2
This work has made use of data from the European Space Agency (ESA) mission {\it Gaia} (\url{https://www.cosmos.esa.int/gaia}), processed by the {\it Gaia} Data Processing and Analysis Consortium (DPAC, \url{https://www.cosmos.esa.int/web/gaia/dpac/consortium}). Funding for the DPAC has been provided by national institutions, in particular the institutions participating in the {\it Gaia} Multilateral Agreement. 
% GMRT
We thank the staff of the GMRT that made these observations possible. GMRT is run by the National Centre for Radio Astrophysics of the Tata Institute of Fundamental Research.
%%%%%%%%%%%%%%%%%%%%%%%%%%%%%%%%%%%%%%%%%%%%%%%%%%



%\bibliographystyle{mnras}

% My notes
% (1) <= 8 authors, list all of their names (Surname X.~Y.~Z.). 
% (2) > 8 authors, only list the first author and use ", et al.," for the rest authors
% (3) Remove "&"  between authors in case of two and multiple author paper in the references. 
% (4) Use "&" between the last two authors in case of citing two or three-author paper. No ","
% (5) Insert a space between initials with "~"

%%%%%%%%%%%%%%%%%%%%%%%%%%%%%%%%%%%%%%%%%%%%%%%%%%

%%%%%%%%%%%%%%%%% APPENDICES %%%%%%%%%%%%%%%%%%%%%

\appendix



%%%%%%%%%%%%%%%%%%%%%%%%%%%%%%%%%%%%%%%%%%%%%%%%%%

% Don't change these lines
\bsp	% typesetting comment
\label{lastpage}
\end{document}